\documentclass[12pt]{article}

\usepackage{color}
\definecolor{darkblue}{rgb}{0.1,0.1,.7}
\usepackage[colorlinks, linkcolor=darkblue, citecolor=darkblue, urlcolor=darkblue, linktocpage]{hyperref}
\usepackage[]{amsmath}
\usepackage[]{graphicx}
\usepackage[]{latexsym}
\usepackage{slashed,graphicx,color,amsmath,amssymb}
\usepackage[mathscr]{eucal}
\usepackage{mathrsfs}
\usepackage{geometry}
\usepackage[margin=10pt,font=small,labelfont=bf]{caption}
\usepackage{amscd}
\usepackage{bm}
\usepackage{xcolor}
\usepackage{upgreek}
\usepackage[square, comma, sort&compress,numbers]{natbib}
\usepackage[all,cmtip]{xy}
\usepackage[color=cyan!30!white,linecolor=red,textsize=footnotesize]{todonotes}
\numberwithin{equation}{section}
\hypersetup{
	pdfstartview={FitH},    
	pdftitle={TT deformed partition functions},    
	pdfauthor={Shouvik Datta, Yunfeng Jiang},     
	colorlinks=true,       
	linkcolor=blue,          
	citecolor=blue!59!black! ,        
	filecolor=magenta,      
	urlcolor=blue           
}

\newcommand{\tr}{\mathrm{Tr}\,}

\def\ttb{T{\bar{T}}}
\def\btau{{\bar{\tau}}}
\def\Z{\mathcal{Z}}
\def\pd{\partial}
\def\D{\mathsf{D}}
\def\DD#1{{\mathscr{D}}^{(#1)}}
\def\E{{\mathcal{E}}}

 \makeatletter
 \g@addto@macro\bfseries{\boldmath}
 \makeatother

\begin{document}
\vspace*{-.6in} \thispagestyle{empty}
\vspace{.2in} {\Large
\begin{center}
{\bf $T\bar{T}$ deformed partition functions}
\end{center}}
\vspace{.2in}
\begin{center}
Shouvik Datta \& Yunfeng Jiang
\\
\vspace{.3in}
\small{\textit{Institut f{\"u}r Theoretische Physik,
ETH Z{\"u}rich}},\\
\small{\textit{
Wolfgang Pauli Strasse 27,
CH-8093 Z{\"u}rich, Switzerland.}
}
\\
\vspace{.3in}
\small{\texttt{$\mathtt{\lbrace}$shouvik,jiangyu$\mathtt{\rbrace}$@itp.phys.ethz.ch}
}

\end{center}

\vspace{.3in}

\begin{abstract}
\normalsize{We demonstrate the presence of modular properties in partition functions of $T\bar{T}$ deformed conformal field theories. These properties are verified explicitly for the deformed free boson. The modular features facilitate a derivation of the asymptotic density of states in these theories, which turns out to interpolate between Cardy and Hagedorn behaviours. We also point out a sub-sector of the spectrum that remains undeformed under the $T\bar{T}$ flow. Finally, we comment on the deformation of the CFT vacuum character and its implications for the holographic dual.}
\end{abstract}

\vskip 1cm \hspace{0.7cm}

\newpage

\setcounter{page}{1}
\begingroup
\hypersetup{linkcolor=black}
\tableofcontents
\endgroup

\section{Introduction}
\label{sec:1}
The search for interacting and yet exactly solvable examples of quantum field theories has always been a major theme in theoretical physics. The best known examples of such theories contain integrable structures or possess special symmetries. The study of these theories provide a wealth of techniques which underlies a large part of our understanding about the microscopic world and collective phenomena.

Conformal field theories form a broad class of solvable theories. One of their best known uses is the description of fixed points of RG flows. In fact, the space of quantum field theories can be classified as fixed points which are conformally invariant and their deformations. While there has been extensive studies on relevant and marginal deformations, the study of irrelevant deformations remains largely unexplored. Irrelevant deformations, by definition, do not affect the IR behavior but will generally affect UV physics. It is challenging to study irrelevant deformations due to both technical and conceptual difficulties. In general, it is unclear whether an irrelevant deformation will lead to a well-defined theory in the UV limit.\par

Recently, a very interesting kind of irrelevant deformation called $T\bar{T}$-deformation has been proposed \cite{Smirnov:2016lqw,Cavaglia:2016oda} and it has attracted growing attention. The $T\bar{T}$ deformation can be defined for any quantum field theory in 1+1 dimensions (see also \cite{Cardy:2018sdv,Taylor:2018xcy} for higher dimensional proposals). It defines a continuous family of theories along a trajectory in the field theory space whose tangent vector at each point is defined as the $T\bar{T}$ operator, which is nothing but $-\frac{1}{2}\det T_{\mu\nu}$£¬ where $T_{\mu\nu}$ is the stress-energy tensor of the theory. This deformation forms a special class RG trajectories that flow `up' from the IR to the UV \cite{Giveon:2017myj}.
\par

The $T\bar{T}$ deformation has a number of remarkable features which make it interesting. To start with, although the deformation leads to rather complicated and usually non-local Lagrangians \cite{Cavaglia:2016oda,Bonelli:2018kik}, the deformed spectrum can be determined non-perturbatively and takes a remarkably compact form. What's more interesting is that this fact can be derived from several different points of view.\par

The original interest in this kind of deformation comes from the study of integrable field theories. It is shown in \cite{Smirnov:2016lqw,Cavaglia:2016oda} that the $T\bar{T}$ operator is the lowest dimensional member of a family of irrelevant operators which trigger \emph{integrable} deformations for integrable quantum field theories. The deformation is integrable in the sense that it preserves an infinite number of conserved charges along the flow. At the level of the factorized $S$-matrix, the $T\bar{T}$ deformation simply modifies the $S$-matrix by multiplying a CDD factor. This makes it possible to adapt many powerful techniques such as Bethe ansatz and form factor bootstrap approach to study the deformed theories. By using the thermodynamic Bethe ansatz one can see easily that for integrable field theories, the spectrum deforms in a simple way.
In fact, an alternative way to find the deformed spectrum is also proposed in \cite{Smirnov:2016lqw,Cavaglia:2016oda} where the authors realized the deformed energy level satisfies the generalized Burgers' equation. This relation is based on an earlier result on the expectation value of $T\bar{T}$ by Zamolodchikov \cite{Zamolodchikov:2004ce}. The Burgers' equation is well-known in fluid dynamics and can be solved readily in this case which gives the deformed spectrum.\par

Recently, Cardy \cite{Cardy:2018sdv} proposed a different point of view on the $T\bar{T}$ deformation. In this approach, the $T\bar{T}$ deformation is interpreted as coupling the field theory to uniform stochastic gravity. The fact that the $T\bar{T}$ deformation is solvable is because the action of the stochastic gravity is a total derivative and has little to do with integrability.
At the same time, yet another point of view is proposed in \cite{Dubovsky:2017cnj,Dubovsky:2018bmo}. The $T\bar{T}$ deformation is also interpreted as coupling the theory to a gravity theory, this time a more specific 2d gravity theory which is Jackiw-Teitelboim gravity.
These two approaches, which reproduce the spectrum obtained from integrability, also strongly hint that under the $T\bar{T}$ deformation, the theory eventually becomes non-local and has features of gravitational theories. These features have been observed already in the previous studies. In particular, it is found that by deforming the Lagrangian of a free boson theory, one ends up with the Nambu-Goto action, which describes a string theory. Therefore, the $T\bar{T}$ deformation might also provide a new window of investigating toy models of quantum gravity, which makes it even more intriguing.\par

Let us also mention that there are also interesting proposals on the worldsheet \cite{Giveon:2017nie,Giveon:2017myj} which share the same feature and can be seen as a holographic description of the $T\bar{T}$ deformation. When the coupling constant takes a negative sign, a holographic dual of the $T\bar{T}$ deformation has been proposed to be AdS$_3$ with a radial cut-off \cite{McGough:2016lol}. Other related interesting developments can be found in \cite{Tateo:older,Guica:2017lia,Shyam:2017znq,Asrat:2017tzd,Giribet:2017imm,Kraus:2018xrn,Aharony:2018vux,Bzowski:2018pcy,Bonelli:2018kik,Chakraborty:2018kpr,We'veBeenAskedToCite1,We'veBeenAskedToCite2}.\par

In this paper, we revisit the partition function of $T\bar{T}$ deformed CFT. Our main finding is a modular property of the torus partition function. The torus partition function of CFT is invariant under modular transformations. This fact leads to many highly non-trivial results for CFTs, such as the Cardy's formula for the asymptotic density of states \cite{Cardy:1986ie}. The $T\bar{T}$ deformed CFT is no longer conformal invariant and it is not clear a priori whether there should be well behaved modular properties for the deformed torus partition function at all. In fact, it is a difficult problem to find modular properties of deformed CFT partition functions\footnote{See $e.g.$ \cite{Dijkgraaf:1996iy,Datta:2014zpa,Iles:2014gra,Kaneko,Wang:1997wj} for holomorphic deformations of CFT where perturbation theory gives rise to quasi-modular forms.}.
Surprisingly, as we will show in what follows, the deformed partition function has very simple modular behaviour at each order in the $\ttb$ coupling. This leads to the following modular property
\begin{align}
\label{eq:Z-deform}
\Z(\tau,\bar{\tau}|\lambda)= \Z\left(  \frac{a\tau +b}{c\tau+d} ,  \frac{a\btau+b}{c\btau+d} \  \Bigg| \  \frac{\lambda}{|c\tau+d|}    \right), \nonumber
\end{align}
where, $\lambda$ is the dimensionless perturbative-expansion parameter, $\ell/|\omega_1|$  ($\ell^2$ is the $\ttb$ coupling and $|\omega_1|$ is the radius of the quantization circle).
We demonstrate this for general deformed CFTs using the deformed energy spectrum found in \cite{Smirnov:2016lqw,Cavaglia:2016oda} and also the diffusion equation \cite{Cardy:2018sdv}. We also verify this in detail for the free boson as an example. This result demonstrates that, in addition to the integrable structure, QFTs along the $\ttb$ trajectory also have modular features inherited from the CFT at the IR fixed point. As we can see from (\ref{eq:Z-deform}), the modular transformation leads to rescaling of the parameter $\ell$. Recalling that $\ell$ parameterizes different theories along the trajectory, this indicates that general modular transformations maps the torus partition function from one theory to another along the trajectory of $T\bar{T}$ deformed theories.

Using this modular property, we derive the high energy asymptotics of the density of states in the $T\bar{T}$ deformed CFTs. The density of states interpolates the Cardy like behavior $\rho(E)\sim e^{\sqrt{E}}$ in the IR and a Hagedorn like behavior $\rho(E)\sim e^{E}$ for $E\to\infty$ in the UV. Since the Hagedorn behaviour is a typical density growth for a string theory, this is a sharp signal that as we deform the theory, it finally becomes non-local and behaves like a gravity theory. The transition for the density of states was first noticed in \cite{Giveon:2017myj} by thermodynamic considerations. Here we provide an independent proof and obtain the logarithmic corrections from the modular property of the partition function.\par

We also point out that there is a special sector in the spectrum which remains untouched by the $T\bar{T}$ deformation. The states in this special sector satisfy a BPS like condition although we do not need supersymmetry. This implies that the elliptic genus remains invariant under the $T\bar{T}$ deformation. The nice modular property and the invariant elliptic genus reveal another internal simplicity of the $T\bar{T}$ deformed theories. 

The paper is structured as follows. In Section \ref{sec:modular-general} we provide the general formalism to derive the modular transformation property of the $\ttb$ deformed partition function. This utilises both the explicit form of the deformed energy spectrum as well as the diffusion equation for the $\ttb$ deformed partition function.  Section \ref{sec:bosons} contains the explicit calculations for deformed free bosons which verify the general formalism. We find the asymptotic spectral density in Section \ref{sec:bootstrap}. We point out a protected sector under $\ttb$ deformations in Section \ref{sec:BPS_sector}.
Some implications of the deformed partition function for the conjectured AdS$_3$ dual is given Section \ref{sec:holography}. We conclude in Section \ref{sec:conc}. The appendices contain some technical details on modular forms and their derivatives and higher order expressions.



 \section{Modular properties of the partition function}
\label{sec:modular-general}
In this section, we prove the modular properties for the deformed partition function (\ref{eq:Z-deform}). We first show that the modular properties can be seen explicitly from a perturbative expansion of the deformed partition function. Then we give a proof by mathematical induction based on the diffusion equation satisfied by the deformed partition function.

\subsection{Perturbation theory for the partition function}
We shall study $\ttb$ deformations of conformal field theories on the torus. Consider a torus with complex periods $\omega_1$ and $\omega_2$. The partition function of a CFT on this genus-one surface is
\begin{align}
Z(\tau,\bar{\tau}) = \tr [q^{L_0-\frac{c}{24}}\bar{q}^{\bar{L}_0-\frac{c}{24}}] = \sum_n e^{2\pi i \tau_1 k_n -  \tau_2 |\omega_1| E_n},
\end{align}
where $q=e^{2\pi i\tau}$ and the modular parameter is $\tau=\omega_2/\omega_1= \tau_1+i\tau_2$. The Hamiltonian of the CFT is $\hat{H}=\tfrac{2\pi}{|\omega_1|}(L_0 +\bar{L}_0-c/12)$, \textit{i.e.,} the theory is quantized on a circle of radius $|\omega_1|$. The $E_n$ appearing above are the eigenvalues of the $\hat{H}/|\omega_1|$.
Owing to large diffeomorphisms of the torus, the partition function is modular invariant
\begin{align}
Z(\tau,\bar{\tau}) = Z\left( \frac{a\tau+b}{c\tau+d},  \frac{a\btau+b}{c\btau+d}\right).
\end{align}
for $a,b,c,d \, \in \, \mathbb{Z}$ and $ad-bc=1$. For future convenience we also note the action of an element $\gamma$ of the modular group SL$(2,\mathbb{Z})$ on the periods
\begin{align}\label{eq:om-trans}
\begin{pmatrix}
\omega'_2 \\ \omega'_1
\end{pmatrix} = \begin{pmatrix}
a &b \\ c &d
\end{pmatrix}
\begin{pmatrix}
\omega_2 \\ \omega_1
\end{pmatrix}.
\end{align}
Of particular interest is the S-modular transformation $\tau \mapsto -1/\tau$. This modular transformation relates information of the light spectrum of the CFT to high energy asymptotics \cite{Cardy:1986ie,km}.

We shall now consider the $T\bar{T}$ deformation of the partition function\footnote{We use the conventions of \cite{Dubovsky:2018bmo}.}. We give the definition at the level of Lagrangian density. Let us consider a family of deformed theories along a trajectory parameterized by $\ell^2$. 
 At the each point on the trajectory, the Lagrangian density is denoted by $\mathcal{L}^{(\ell^2)}$ and the infinitesimally deformed Lagrangian density is given by
\begin{align}
\mathcal{L}^{(\ell^2+\delta\ell^2)}=\mathcal{L}^{(\ell^2)}+\frac{\delta\ell^2}{2\pi^2}\det T^{(\ell^2)}_{\mu\nu},
\end{align}
where $T_{\mu\nu}^{(\ell^2)}$ is the stress-energy tensor for the theory $\mathcal{L}^{(\ell^2)}$. We want to emphasise that the deformation at each point is triggered by the stress-energy tensor defined at that point and, in general, it is quite different from the stress-energy tensor of the original theory. In the presence of this deformation of the CFT, the spectrum can be solved completely. The deformed energies are
\begin{align}\label{eq:deformed-energies}
\mathcal{E}^{(\ell)}_n = \frac{|\omega_1|}{\ell^2} \left[\sqrt{ 1+\frac{2\ell^2 E_n}{|\omega_1|} +\frac{\ell^4P_n^2}{|\omega_1|^2} } -1  \right],
\end{align}
where $E_n$ and $P_n = 2\pi k_n/|\omega_1|$ are the energies and momenta of the undeformed CFT. The momenta $P_n$ do not change under this deformation. 
The partition function of the deformed theory is then
\begin{align}\label{eq:part1}
\Z(\omega_1,\omega_2,\ell) =   \sum_n e^{  i \tau_1|\omega_1| P_n -  \tau_2 |\omega_1| \mathcal{E}^{(\ell)}_n}.
\end{align}
There are two choices of the $T\bar{T}$ coupling in the literature which correspond to $\ell$ being real or purely imaginary. For the second case (see for example \cite{Cavaglia:2016oda,McGough:2016lol}), the deformed spectrum gets truncated beyond which it becomes imaginary. Our discussion in this section work for both cases.

Plugging in the explicit form of the new spectrum \eqref{eq:deformed-energies} and expanding in powers of the dimensionless parameter $\lambda\equiv\ell/|\omega_1|$, we obtain a perturbative expansion for the partition function
\begin{align}
\Z(\tau,\bar{\tau}|\lambda)=\sum_{m=0}^{\infty} \lambda^{2m} Z_{m}(\tau,\btau) \ .
\end{align}
Note that we have slightly changed the notation from \eqref{eq:part1} but we are referring to the same quantity. The values of $Z_m$ at the first few orders are
\begin{align}
Z_0=&\,\sum_n e^{i\tau_1|\omega_1|P_n-\tau_2 |\omega_1|E_n}, \nonumber\\
Z_1=&\,|\omega_1|^2\sum_n\left(\frac{\tau_2}{2}(E_n^2-P_n^2)\right)e^{i\tau_1|\omega_1|P_n-\tau_2 |\omega_1|E_n},\\\nonumber
Z_2=&\,|\omega_1|^4\sum_n\left(\frac{\tau_2^2}{8}(E_n^2-P_n^2)^2-\frac{\tau_2}{2}(E_n^2-P_n^2)\frac{E_n}{|\omega_1|}\right)e^{i\tau_1|\omega_1|P_n-\tau_2 |\omega_1|E_n},\\\nonumber
Z_3=&\,|\omega_1|^6\sum_n\left(\frac{\tau_2^3}{48}(E_n^2-P_n^2)^3-\frac{\tau_2^2}{4}(E_n^2-P_n^2)^2\frac{E_n}{|\omega_1|}
+\frac{\tau_2}{8}(E_n^2-P_n^2)\frac{5E_n^2-P_n^2}{|\omega_1|^2}\right)e^{i\tau_1|\omega_1|P_n-|\omega_1|E_n}.
\end{align}

The various powers of $E_n$ and $P_n$ appearing at each order in perturbation theory can be converted into actions of derivatives on the CFT partition function, $Z_0=Z(\tau,\btau)=\Z(\tau,\btau|0)$. The simple property
\begin{align}
\label{eq:Zexpansion}
\partial_{\tau_1}^m\partial_{\tau_2}^n Z_0=\left(i |\omega_1| P_n \right)^m\left(-|\omega_1| E_n\right)^n Z_0 ,
\end{align}
leads to the following replacement rule
\begin{align}
E_n\mapsto -\frac{1}{|\omega_1|}\partial_{\tau_2},\qquad P_n\mapsto \frac{1}{i|\omega_1|}\partial_{\tau_1} \ \ .
\end{align}
The derivatives $\pd_{\tau_{1,2}}$ can be easily converted to $\pd_{\tau}$ and $\pd_{\btau}$ using
$
\pd_{\tau_1} = \pd_{\tau} + \pd_{\btau} ,  \  \pd_{\tau_2} = i(\pd_{\tau} - \pd_{\btau} ).
$
Each order in the perturbative expansion $Z_n$ can therefore be related to derivatives of $Z_0$ with respect to the modular parameter (and its conjugate). For example
\begin{align}
\hspace{-.4cm}Z_1=&\,{2\tau_2}\left(\partial_{\tau}\partial_{\btau}\right)Z_0, \label{eq:diff1}\\
\hspace{-.4cm}Z_2=&\,\left[{2\tau_2^2}\left(\partial_{\tau}\partial_{\btau}\right)^2+ {2i\tau_2}\left(\partial_{\tau}-\partial_{\btau}\right)  \left(\partial_{\tau}\partial_{\btau}\right) \right]Z_0, \label{eq:diff2}\\
\hspace{-.4cm}Z_3=&\,\left[ \frac{4\tau_2^3}{3}\left(\partial_{\tau}\partial_{\btau}\right)^3+{4i\tau_2^2} \left(\partial_{\tau}-\partial_{\btau}\right)  \left(\partial_{\tau}\partial_{\btau}\right)^2 -{2\tau_2}  \left( \pd_\tau^2 -3 \pd_\tau\pd_\btau +\pd_\btau^2  \right) \left(\partial_{\tau}\partial_{\btau}\right) \right]Z_0 . \label{eq:diff3}
\end{align}

This shows that, at a formal level, the deformed partition function can be generated from the undeformed one by the action of derivatives
\begin{align}
\Z(\tau,\bar{\tau}|\lambda)=\mathcal{D}_{\tau,\btau}\Z(\tau,\bar{\tau}|0),
\end{align}
where, the `flow operator' $\mathcal{D}_{\tau,\btau}$ is defined non-perturbatively as
\begin{align}
\mathcal{D}_{\tau_1,\tau_2}=:\!\exp\left[-\frac{\tau_2}{\lambda^2}\left(\sqrt{1-2\lambda^2\partial_{\tau_2}-{\lambda^4}\partial_{\tau_1}^2}-1 \right)-\tau_2\partial_{\tau_2}\right]\!: \, ,
\end{align}
where the normal ordering means in the perturbative expansion of the above operator, we put all the operators to the right of factors involving $\tau_2$.

\subsection{Modular properties from perturbation theory}
Let us now investigate the modular properties of $\Z(\tau,\bar{\tau}|\ell)$ order by order in perturbation theory using the formalism of the previous subsection. At the zeroth order, $Z_0$, the partition function is the same as that of a CFT and is modular invariant.

It will turn out to be useful to understand the action of the derivatives $\pd_\tau$ and $\pd_\btau$ on modular functions of various weights.
By using the chain rule, it can be easily seen that the derivative $\pd_\tau$ transforms under modular transformation as
\begin{align}
\pd_{\tau}  = \pd_\tau (\gamma\cdot \tau) \,  \pd_{\gamma\cdot \tau } = \frac{1}{(c\tau+d)^2} \pd_{\gamma\cdot \tau } .
\end{align}
This implies that the action of $\pd_\tau$ on a modular invariant function increases the modular weight by 2. Note that for modular functions of zero weight the resulting function, after differentiation, is still modular. The first order correction contains the factor $\left(\partial_{\tau}\partial_{\btau}\right) Z_0$, this is a modular function of holomorphic and anti-holomorphic weights 2 each.

The situation is subtle when the derivative acts on a modular function of non-zero weight, $f_k(\gamma\cdot\tau)=(c\tau+d)^{k}f_k(\tau)$. We encounter this at the second order \eqref{eq:diff2}, where the derivatives act on a modular function of weight 2. In general, the action of the derivative on a weight $k\neq 0$ modular function does not yield something modular
\begin{align}
\pd_{\gamma\cdot\tau} f_k(\gamma\cdot\tau) = (c\tau+d)^{k+2}\pd_{ \tau} f_k( \tau) + k c (c\tau+d)^{k+1}  f_k( \tau) .
\end{align}
There are a number of ways to construct differential operators which preserve modularity. A well-known one is the Ramanujan-Serre derivative which corrects the second term above by the anomalous transformation of $E_2(\tau)$. However, the operator   which turns out to be useful for present purposes is (see equation~(55) in \cite{Zagier})
\begin{align}\label{eq:covD}
\D^{(k)}_\tau f_k(\tau) \equiv \pd_\tau f_k(\tau) - \frac{i k}{2 \tau_2}f_k(\tau).
\end{align}
The resulting function is a non-holomorphic modular form of weight $k+2$. The strategy now is to verify whether combinations of derivatives appearing in equations \eqref{eq:diff2}, \eqref{eq:diff3} and at higher orders can be rewritten fully in terms of $\D^{(k)}_{\tau}$ and $\D^{(k)}_{\btau}$. If this holds, a modular structure can be found at each order in perturbation theory. Note that the second and higher orders in perturbation theory are not guaranteed to be modular since the $\ttb$ deformation involves an infinite number of terms and is not $T$ and $\bar{T}$ of the undeformed CFT.

It can be checked perturbatively that the above expectation is indeed true, \textit{i.e.}, the combinations of derivatives appearing in \eqref{eq:diff2}, \eqref{eq:diff3} and at higher orders are fully expressible in terms of the derivatives \eqref{eq:covD}. We define the following operator to make the expressions compact
\begin{align}
\mathscr{{D}}^{(j)} = \prod_{m=0}^{j} {\mathsf{D}}^{(2m)}_\tau {\mathsf{D}}^{(2m)}_\btau
\end{align}
Acting this operator on a modular invariant object converts it to a non-holomorphic modular object of weight $4(j+1)$.
The perturbative corrections to the partition function now take the following form
\begin{align}
Z_1=&\,{2\tau_2} \, \mathscr{{D}}^{(0)}Z_0, \label{eq:ndiff1}\\
Z_2=&\, {2\tau_2^2} \, \left[\mathscr{{D}}^{(1)} - \frac{1}{2\tau_2^2}\mathscr{{D}}^{(0)} \right]
Z_0, \label{eq:ndiff2}\\
Z_3=&\, \frac{4\tau_2^3}{3}\left[ \DD{2} - \frac{2}{\tau_2^2} \DD{1} + \frac{{3}}{4\tau_2^4} \DD{0} \right]Z_0 , \label{eq:ndiff3} \\
Z_4=&\, \frac{2\tau_2^4}{3}\left[ \DD{3} - \frac{5}{\tau_2^2} \DD{2} + \frac{{27}}{4\tau_2^4} \DD{1} -\frac{{9}}{4\tau_2^6} \DD{0} \right]Z_0 . \label{eq:ndiff4}
\end{align}
Higher orders can also be computed algorithmically and we provide the results in appendix~\ref{sec:diffOp}. This suggests the following structure for the $p$th order partition function
\begin{align}
\label{eq:explicitZp}
Z_p =   \sum_{m=0}^{p-1} {a_{p,m} \over \tau_2 ^ {p-2(m+1)} }\DD{m} Z_0
\end{align}
for some rational fractions $a_{p,m}$. Recall that $\tau_2$ is non-holomorphic and modular with weight 2,
$
\text{Im}(-1/\tau) = |c\tau+d|^{-2} \,  \text{Im}(\tau)
$ and $\DD{m}Z_0$ has modular weight $4(m+1)$. This implies that each term within the summation symbol above are modular and transform with weight $2p$
\begin{align}
\label{eq:Zptransform}
Z_p(\gamma\cdot\tau,\gamma\cdot\bar{\tau})={|c\tau+d|} ^{2p}  Z_p(\tau,\bar{\tau}).
\end{align}
That is $Z_p$ is a non-holomorphic modular form of weight $2p$. This is a non-trivial feature of the $\ttb$ deformation.

\subsection{Perturbation theory and the diffusion equation}
From the perturbative expansion, we have strong evidence that each $Z_p$ is a non-holomorphic modular form of weight $2p$, equation \eqref{eq:Zptransform}. In this subsection, we prove this modular property at arbitrary orders in perturbation theory by using the diffusion equation derived in \cite{Cardy:2018sdv}\footnote{We are grateful to Alex Maloney \cite{am} and John Cardy for discussions which led to this subsection.}. For the torus, the deformed partition function satisfies the following diffusion relation, in the convention of \cite{Cardy:2018sdv}
\begin{align}
\label{eq:diffusioneq}
\partial_t\mathcal{Z}=(\partial_{L_1}\partial_{L'_2}-\partial_{L_2}\partial_{L'_1})\mathcal{Z}
-\frac{1}{A}\left[(L_1\partial_{L_1}+L'_1\partial_{L'_1})+(L_2\partial_{L_2}+L'_2\partial_{L'_2})\right]\mathcal{Z},
\end{align}
where
\begin{align}
\label{eq:change}
\omega_1=L_1+iL_2,\qquad \omega_2=L'_1+iL'_2,\qquad\tau=\frac{\omega_2}{\omega_1}.
\end{align}
The relation between the couplings is $t=-\ell^2/2$ and the area $A$ is $L_1L'_2-L_2L'_1$. We note that modular invariance of the deformed partition function is guaranteed from the flow equation. This is because the differential operator and the boundary condition ($\mathcal{Z}$ at $t=0$) are modular invariant.

Plugging the   the expansion (\ref{eq:Zexpansion}) into the diffusion equation (\ref{eq:diffusioneq}) and comparing the coefficients\footnote{It is important to compare coefficients of $\ell^{2p}$ and not $\lambda^{2p}$. The derivatives $\pd_{L_{i}}$ on $\lambda=\ell/|\omega_1|$ are non-zero because of \eqref{eq:change}.} of $\ell^{2p}$, we obtain a relation between $Z_{p+1}$ and $Z_p$
\begin{align}
\label{eq:recursionZp}
Z_{p+1}=&\,\frac{|\omega_1|^2}{2(p+1)}(\partial_{L_1}\partial_{L'_2}-\partial_{L_2}\partial_{L'_1})Z_p
+\frac{p}{p+1}(L_2\partial_{L'_1}-L_1\partial_{L'_2})Z_p\\\nonumber
&\,-\frac{|\omega_1|^2}{2A(p+1)}(L_1\partial_{L_1}+L'_1\partial_{L'_1}+L_2\partial_{L_2}+L'_2\partial_{L'_2})Z_p+\frac{p|\omega_1|^2}{A(p+1)}Z_p.
\end{align}
Using the fact that $Z_p=Z_p(\tau,\bar{\tau})$ and the relations (\ref{eq:change}), we can rewrite the recursion relation (\ref{eq:recursionZp}) in a much more compact form
\begin{align}
Z_{p+1}=\frac{1}{p+1}\left(2\tau_2\partial_{\tau}\partial_{\bar{\tau}}+ip(\partial_{\tau}-\partial_{\bar{\tau}})
-\frac{p}{\tau_2}\right)Z_p
\end{align}
where we have used the fact that $A/|\omega_1|^2=\tau_2$. We have checked that the recursive use of this reproduces \eqref{eq:diff1}-\eqref{eq:diff3}. Written in terms of the covariant modular derivatives defined in (\ref{eq:covD}), we simply have
\begin{align}
\label{eq:Zrecurs}
Z_{p+1}=\frac{2\tau_2}{p+1}\left(\D_{\tau}^{(p)}\D_{\bar{\tau}}^{(p)}-\frac{{p(p+1)}}{4\tau_2^2}\right)Z_p.
\end{align}
We now perform mathematical induction. Assuming that $Z_p$ is a non-holomorphic modular form of weight $2p$ (\ref{eq:Zptransform}), then the above equation (\ref{eq:Zrecurs}) clearly shows that $Z_{p+1}$ is a non-holomorphic modular form of weight $2(p+1)$. This completes the proof.

It is worthwhile to note that we get \textit{modular} forms at each order in perturbation theory.
Usually upon turning coupling/chemical potentials in CFTs, conformal perturbation theory leads to \textit{quasi-modular} forms at each order -- at least for the case of holomorphic irrelevant deformations \cite{Dijkgraaf:1996iy,Datta:2014zpa,Iles:2014gra}. Furthermore, because of this quasi-modularity it is often hard to find modular transformation properties of charged partition functions. In case of  a non-zero $U(1)$ chemical potential, the partition function transforms covariantly under modular transformations. For the $\ttb$ deformation, each order in conformal perturbation theory gives a modular function and, therefore, the full partition is modular invariant.

\subsection{Deformed partition function }

Let us now return to the full partition function
\begin{align}
\Z(\tau,\bar{\tau}|\lambda)=\sum_{p=0}^{\infty}  \lambda^{2p}\, Z_p(\tau,\btau)=\sum_{p=0}^{\infty}  \lambda^{2p} \ \sum_{m=0}^{p-1} {a_{p,m} \over \tau_2 ^ {p-2(m+1)} }\DD{m} Z_0 .
\end{align}
The modular transformation of the dimensionless coupling $\lambda$ can be uniquely fixed by demanding that each order in conformal perturbation theory transforms uniformly. Therefore, we arrive at the following modular property for the partition function
\begin{align}\label{eq:mod-trans-Z}
\Z(\tau,\bar{\tau}|\lambda)= \Z\left(  \frac{a\tau +b}{c\tau+d} ,  \frac{a\btau+b}{c\btau+d} \  \Bigg| \  \frac{\lambda}{|c\tau+d|}   \right).
\end{align}
This is the key result of this paper. The transformation of $\lambda$ is analogous to the elliptic variable of a torus, although it is non-holomorphic\footnote{This statement has been suggested by A.~Maloney \cite{am}.}.

It is worthwhile to note how the $\ttb$ coupling $\ell= \lambda/|\omega_1|$ transforms. In order to see this, we recall from \eqref{eq:om-trans} that
\begin{align}\label{eq:om2}
|\gamma\cdot\omega_1| = |c\tau+d| \times |\omega_1|.
\end{align}
This along with \eqref{eq:mod-trans-Z} implies that $\ell$ remains invariant under modular transformations.
%
%
%
%
%

\section{An example : the free boson}
\label{sec:bosons}

\subsection{Deforming the free boson theory}
In this section, we consider the deformed $c=1$ free boson in detail as an explicit example for our general discussion in the previous section. The advantage is that in this case we know the undeformed partition function and the perturbative corrections $Z_m$ explicitly in closed form.
Although this is one of the simplest CFTs, the $\ttb$ deformation is non-trivial. Let us denote the free boson Lagrangian density as
\begin{align}
\mathcal{L}_{\text{free-boson}}=\partial\phi(z,\bar{z})\bar{\partial}\phi(z,\bar{z}).
\end{align}
The undeformed partition function is given by
\begin{align}
\Z(\tau,\btau|0)=\frac{1}{\sqrt{\tau_2}\, \eta(\tau)\eta(\bar{\tau})},
\end{align}
where $\eta(\tau)$ is the Dedekind eta function. The factor involving the Dedekind etas comes from oscillators, whilst the $1/\sqrt{\tau_2}$ arises from the continuous zero modes.

The $\ttb$ deformed Lagrangian is given by \cite{Cavaglia:2016oda}
\begin{align}
\mathcal{L}^{(\ell^2)}=\frac{1}{\ell^2}\left(\sqrt{2\ell^2\,\partial\phi\bar{\partial}\phi+1}-1\right).
\end{align}
The square root part is exactly the same as a Nambu-Goto action for a bosonic string theory in a 3 dimensional flat target space
\begin{align}
\sqrt{2\ell^2\,\partial\phi\bar{\partial}\phi+1}=\sqrt{\det\left(\partial_{\alpha}X\cdot\partial_{\beta}X\right)} \ ,
\end{align}
in the static gauge,
$
X^1=x,\, X^2=y,\, X^3=\frac{\ell}{\sqrt{2}}\phi
$.
Since this is a theory of strings, it demonstrates non-local features upon turning on this deformation. In this gauge, $\ell$ directly maps on to the length of the string, $\ell_s=\sqrt{\alpha'}$. Clearly, in the point-particle limit of the string,  $\ell_s\to0$, the theory is local, whilst non-local features emerge when $\ell_s>0$. Another way to rephrase this is that increasing the $\ttb$ coupling takes us from the tensile towards the tensionless regime of strings.  In a sense, the perturbative expansion which follows is in the same vein of the $\alpha'$ expansion of string perturbation theory at one-loop ($i.e.,$ at a fixed genus, $g=1$). Computations similar to the one which follows  have been performed in the context of the relation  of effective strings to lattice  gauge theories   \cite{Dietz,Luscher:2004ib,Billo:2005iv,Billo:2006zg,Tateo:older,Billo:2011fd}.


\par

\subsection{Perturbation theory}

We shall now investigate the deformed partition function perturbatively. In particular, we shall use the equations \eqref{eq:diff1}, \eqref{eq:diff2}, \eqref{eq:diff3} and their higher order analogues to find the perturbative corrections. The resulting expressions can be written down in closed form in terms of Eisenstein series. The modular properties of the perturbation series will turn out to be exactly as predicted by the general analysis of the previous section.

\subsubsection*{First order}
The first order correction is given by \eqref{eq:diff1}.
In order to proceed, we use the relation of the derivative of $\eta(\tau)$ with the Eisenstein series $E_2(\tau)$ \eqref{eq:deta2}.
The relations for the antiholomorphic part is exactly the same. 
Using the equation (\ref{eq:deta2}) and reorganizing terms, we find ($\text{Im}\,\bar{\tau}=-\tau_2$)
\begin{align}
Z_1=\frac{2Z_0}{(12)^2\tau_2}\left[-\left(E_2(\tau)-\frac{3}{\pi\,\text{Im}\,\tau}\right)
\left(E_2(\bar{\tau})-\frac{3}{\pi\, \text{Im}\,\bar{\tau}}\right)\pi^2\tau_2^2+18\right].
\end{align}
It is known that $E_2(\tau)$ is not modular but quasi-modular. However, the precise combination
\begin{align}\label{eq:tE2}
\tilde{E}_2(\tau)=E_2(\tau)-\frac{3}{\pi\, \text{Im}\,\tau}
\end{align}
is a modular form of holomorphic weight 2, albeit the function itself being non-holomorphic, namely $\tilde{E}_2(-1/\tau)=\tau^2\tilde{E}_2(\tau)$ \cite{Dijkgraaf:1996iy,Datta:2014zpa}. We can write $Z_1$ compactly as
\begin{align}
\label{eq:Z1boson}
Z_1= {2\tau_2} \left[-{\pi^2\over 144}|\tilde{E}_2|^2+{1\over 8 \tau_2^2}\right]Z_0
\end{align}
The structure of this expression is precisely of the form \eqref{eq:ndiff1}. The quantity within square brackets along with $Z_0$ has modular weight 4 and can be identified with $\DD{0}Z_0$ in \eqref{eq:ndiff1}. $Z_1$ as a whole is a non-holomorphic modular form of weight 2.  It is rather remarkable that the combination \eqref{eq:tE2} precisely appears in the expression to yield a modular function. However, this feature is not so surprising at the first order since the $\tau$-derivative on a modular invariant function yields a modular function of weight 2. As mentioned in the previous section modularity at higher orders is not guaranteed since derivatives of non-zero weight modular functions are not modular. 

\subsubsection*{Second order}
It is straightforward to calculate the next orders in perturbation theory. In order to get closed form expressions we use the Ramanujan identities \eqref{eq:dtau} which provide the derivatives of Eisenstein series. These identities also allow us to simplify higher order derivatives of $\eta(\tau)$.
The second order correction is given by
\begin{align}
Z_2= {2\tau_2^2} \left[\left({\pi^4\over 20736}\left|\tilde{E}_2^2-2E_4\right|^2 +{1\over 32\tau_2^4}\right) -\frac{1}{2\tau_2^2}\left(-{\pi^2\over 144}\left|\tilde{E}_2\right|^2+{1\over 8\tau_2^2}\right)\right]Z_0
\end{align}
Once again this is exactly of the form \eqref{eq:ndiff2} from the general analysis. The above expression is reorganized on purpose to show the appearance of $\DD{0}Z_0$ and $\DD{1}Z_0$. The quantity in square brackets is a non-holomorphic modular function of weight $8$ and $Z_2$ is a non-holomorphic modular form of weight 4 as expected. This is the first non-trivial order at which we see a modular function as opposed to a quasi-modular one.

\subsubsection*{Higher orders}
The third and fourth order corrections can be analogously computed as
\begin{align}
Z_3=&\,\frac{2\tau_2^3Z_0}{(12)^6}\left[-\frac{2\pi^6}{3}\left|3\tilde{E}_2^3-18\tilde{E}_2E_4+16E_6\right|^2\, -{324\pi^4\over \tau_2^2}\left|\tilde{E}_2^2
-2E_4\right|^2\,\right.\\\nonumber
&\qquad\qquad\qquad\left.-{11664\pi^2 \over \tau_2^4}\left|\tilde{E}_2\right|^2\,
+{69984\over\tau_2^6}  \right]\\
Z_4=&\,\frac{2\tau_2^4 Z_0}{(12)^8}\left[\frac{\pi^8 }{3}\left|15\tilde{E}_2^4-180\tilde{E}_2^2E_4-156E_4^2+320\tilde{E}_2E_6\right|^2\,\right.\\\nonumber
&\qquad\qquad\qquad+{480\pi^6\over\tau_2^2}\left|3\tilde{E}_2^3-18\tilde{E}_2E_4+16E_6\right|^2\,+
{11660\pi^4\over\tau_2^4}\left|\tilde{E}_2^2-2E_4\right|^2\,\\\nonumber
&\qquad\qquad\qquad\left.+
{2799360\pi^2\over\tau_2^6}\left|\tilde{E}_2\right|^2\,-{12597120\over\tau_2^8}\right]
\end{align}
The quantities within square brackets are clearly non-holomorphic modular functions of weights 12 and 16 respectively; these combined with the $\tau_2$ factors result in modular weights of 6 and 8 for the full expressions. We have computed $Z_{n}$ up to $n\sim 20$ and confirmed that the modular structure  is consistent with the analysis of the previous section. (The next few orders can be found in Appendix \ref{sec:diffOp}.) Therefore, this serves as a verification of the modular property \eqref{eq:mod-trans-Z}.

\section{Modular bootstrap and asymptotic density of states}
\label{sec:bootstrap}
In this section we shall utilize the modular properties of the partition function to derive an asymptotic formula for density of states in the $\ttb$ deformed theories. This analysis is analogous to the derivation of Cardy's formula for CFT \cite{Cardy:1986ie}. The density of states for the $\ttb$ deformed CFTs have been derived previously in \cite{Giveon:2017myj,McGough:2016lol} using the deformed spectrum and Cardy's formula.  In what follows, we shall provide an independent derivation from the modular property \eqref{eq:mod-trans-Z}. This will also enable us to obtain the logarithmic corrections to the entropy, which will play a role in finding the behaviour of the partition function in the UV regime.

 We shall work with $\ell$ being real in this section, for which the spectrum does not become imaginary or truncated at high energies.
The S-modular property for the partition function is simply
\begin{align}
\Z(\tau,\btau|\lambda) = \Z\left(-{1\over \tau}, -{1\over \btau}\bigg|\ \frac{\lambda}{|\tau|}\right) .
\end{align}
For a rectangular torus with a purely imaginary modular parameter, $\omega_1=R, \ \omega_2=i\beta, \ \tau=i \beta/R$, this is  simply
$
\Z(R,i\beta,\ell) = \Z(iR,\beta, \ell)
$ in the notation of \eqref{eq:part1}.
For $c>0$ theories the ground state in the $\ttb$ deformed theory is the deformed vacuum of the CFT \footnote{A potential concern here may be level crossings under the $\ttb$ deformation. Fortunately, this does not occur if the coupling $\ell$ is real.}, $i.e.,$ equation \eqref{eq:deformed-energies} with $E_0 = -c/12$ and $P_0=0$. Therefore, the low-temperature ($\beta \to \infty$) expansion is
\begin{align}
\Z(R,i\beta,\ell) \ \simeq \ e^{-\beta \ell^{-2} \left(\sqrt{R^2- {c\ell^2 \over 6} }-R\right)} =   e^{-{\beta R\over\ell^{2}} \left(\sqrt{1- {c\ell^2 \over 6R^2} }-1\right)} .
\end{align}
The S-modular transform ($R\leftrightarrow \beta$) of this yields the high temperature behaviour. The partition function can also be written as an integral over the spectrum by introducing the spectral density $\rho(\E)= \sum_n \delta(\E-\E_n^{\ell})$.
\begin{align}
\Z(iR,\beta, \ell)\ \simeq \ e^{-{\beta R \over \ell^{2}} \left(\sqrt{1- {c\ell^2 \over 6\beta^2} }-1\right)} = \int_{\E^(\ell)_0}^\infty \rho(\E) e^{-\beta \E} d\E .
\end{align}
The condition $c\ell^2 \leq 6 \beta^2$ needs to be satisfied in order to ensure reality of the expression above. Since $\beta$ is small at high temperatures, the analysis of this section has a restricted regime of validity.
$\rho(\E)$ can be extracted by the inverse Laplace transform.
\begin{align}
\rho(\E) = \oint d\beta \exp \Bigg\lbrace \beta \E  - \frac{\beta  R}{\ell^2} { \left(\sqrt{1-\frac{c \ell^2}{6 \beta ^2}}-1\right)}\Bigg\rbrace.
\end{align}
The integral above is dominated by a saddle at
\begin{align}
\beta_* =  \frac{1}{\sqrt{{12\over c}\tfrac{\E}{R}}} \ \frac{ (1+ \tfrac{\ell^2 \E}{R})}{{(1+\tfrac{\ell^2 \E}{2R})^{1\over 2}}}.
\end{align}
The first factor is the conventional saddle that appears in the derivation of the Cardy's formula. The second $\ell$ dependent factor is the deformation due to $\ttb$.
Retaining quadratic fluctuations about the saddle point, we are led to the following behaviour for the density of states at high energies
\begin{align}\label{eq:DoS}
\rho(\E) \simeq  \frac{2^{3/4}\pi   R }{\ell^{3/2}} \left(\frac{c}{3}\right)^{1/4} \left[\E(\E+2R/\ell^2)\right]^{-3/4}  \exp \left[  \sqrt{\frac{c \, \E R}{3}{\left(1+{\ell^2 \E\over 2R}\right)} }  \right] .
\end{align}

It can be seen that the density of states exhibits a cross-over between Hagedorn and Cardy regimes.
In the UV regime ($R$ being small), $\ell^2 \mathcal{E} \gg 2R$ we get the Hagedorn behaviour
\begin{align}
\rho(\E) \simeq \ \mathcal{N}_H \  \E^{-3/2}\exp \left[  \sqrt{\frac{c }{6} }\,  \ell \, \E  \right].
\end{align}
Defining the Hagedorn temperature as $\beta_H =  \ell  \sqrt{{c }/{6} }$, the scaling of the partition function with respect to temperature in this regime can be found
\begin{align}
\Z(\beta,\ell) \ \simeq \ \int_{\E^\ell_0}^{\infty} d \E \, \E^{-3/2} \exp \left[ (\beta_H - \beta) \E  \right] \ \simeq \  (\beta-\beta_H)^{-5/2} \  \Gamma(\tfrac{5}{2},\E^{\ell}_0(\beta-\beta_H)).
\end{align}
Here, $\Gamma(a,b)$ is the incomplete gamma function.
In the IR regime $\ell^2 E \ll 2R$,  we recover the standard Cardy growth for a CFT
\begin{align}
\rho(\E)  \simeq \ \mathcal{N}_C \  \E^{-3/4}\exp \left[  \sqrt{\frac{c R }{3} \E} \,\right].
\end{align}

It is worthwhile noting that, although \eqref{eq:DoS} has a restricted regime of validity it still holds true beyond this regime $i.e.$ the $\ttb$ coupling, $\ell$, need not be small. We have checked this explicitly using the spectrum of the deformed free boson theory.

\section{A BPS sector}
\label{sec:BPS_sector}
In this section, we point out an interesting sector in the spectrum that is invariant under the $T\bar{T}$ deformation. This can be easily seen from the deformed energy spectrum (\ref{eq:deformed-energies}). If we take $P_n=E_n$, we have
\begin{align}
\mathcal{E}_n^{(\ell)}=\frac{|\omega_1|}{\ell^2}\left[\sqrt{1+\frac{2E_n\ell^2}{|\omega_1|}+\frac{\ell^4 E_n^2}{|\omega_1|^2}}-1\right]=E_n=P_n
\end{align}
which is invariant. Recall that in the undeformed CFT energy and momentum are related to the conformal dimensions $(\Delta_n,\bar{\Delta}_n)$ as
\begin{align}
E_n=\Delta_n+\bar{\Delta}_n-\frac{c}{12},\qquad P_n=\Delta_n-\bar{\Delta}_n.
\end{align}
The condition $E_n=P_n$ implies
\begin{align}\label{eq:Ramond-gr}
\bar{\Delta}_n - \frac{c}{24} =0 .
\end{align}
In theories with supersymmetry, the above condition is satisfied by Ramond ground states. However, this is a BPS condition which does not necessarily rely on supersymmetry. From the bulk point of view, these states correspond to extremal black holes. This observation reveals another intrinsic simplicity of the $T\bar{T}$ deformation.

For theories with supersymmetry, a  direct consequence of this is that the elliptic genus is conserved along the $\ttb$ flow. This occurs because states obeying \eqref{eq:Ramond-gr} are annhiliated by $\bar{T}$ of the $\ell^2 \ttb$ coupling.
 The elliptic genus is defined as\footnote{We set the chemical potential to zero as in the initial definition \cite{Witten:1986bf} which also applies to non-supersymmetric theories. Alternatively, we could have defined the elliptic genus with a $(-1)^{F_L}$ insertion, which would have contributions only from states satisfying $E_n = -P_n$. }
\begin{align}\label{eq:ell-gen}
\chi(\tau) = \tr_{\text{RR}} \left[(-1)^{F_R} e^{  i \tau_1  \hat P -  \tau_2  \hat{H}^{(\ell)}}\right]. 
\end{align}
Here, $\hat{H}^{(\ell)}$ is the deformed Hamiltonian, $F_R$ is the right-moving fermion number and the factor of $(-1)^{F_R}$  imposes a Bose-Fermi cancellation to restrict the sum only over Ramond ground states, $i.e.,$ states obeying the condition \eqref{eq:Ramond-gr} are the only ones that will survive the trace. This renders the elliptic genus to be independent of $\btau$. Since the sum is over states of the kind \eqref{eq:Ramond-gr} which remain undeformed by $\ttb$, the elliptic genus remains invariant. Therefore, for supersymmetric theories, the quantity \eqref{eq:ell-gen} can  be potentially used to classify various $\ttb$ trajectories.

It would be intriguing to investigate whether this sector remains protected under more general  integrable deformations by $X_s$ corresponding to conserved higher spin currents $\lbrace T_{s+1}, $ $ \bar{T}_{s+1} \rbrace$ \cite{Smirnov:2016lqw}. These take the form
\begin{equation*}
X_s = T_{s+1}\bar{T}_{s+1} - \Theta_{s-1}\bar{\Theta}_{s-1}. 
\end{equation*}
where, $\lbrace \Theta_{s-1}, \bar{\Theta}_{s-1}\rbrace$ are the traces.
This would require knowledge of the deformed spectrum in these cases.

%
%

\section{Comments on holography}
\label{sec:holography}
A holographic dual for $\ttb$ deformed CFTs, for $\ell$ being purely imaginary,  has been proposed in \cite{McGough:2016lol,Taylor:2018xcy} in terms of $AdS_3$ with a finite radial cut-off \cite{Brattan:2011my}. This intriguing proposal reproduces several features of the deformation, including the spectrum, thermodynamics and signal propagation speeds. Furthermore, correlation functions of stress-tensors can be reproduced in this cut-off geometry \cite{Kraus:2018xrn,Aharony:2018vux}.

The formalism developed in Section \ref{sec:modular-general} can be used to find how one-loop determinants of various fields in the bulk would get modified by the cut-off geometry. In particular, the deformation of the vacuum character of the CFT can be found using the differential operators ($e.g.$ \eqref{eq:diff1}, \eqref{eq:diff2} and \eqref{eq:diff3}). It is known that the vacuum character is reproduced by the one-loop determinant of the bulk massless gravitons \cite{Giombi:2008vd,Maloney:2007ud}. Therefore, if the proposal of \cite{McGough:2016lol} is true at one-loop it should agree with the deformed vacuum character. The modified product of the holomorphic and non-holomorphic vacuum character is given by
\begin{align}\label{eq:vac}
&Z_{\rm vac}(\tau,\btau|\ell) = \left|\frac{q^{-\frac{c-1}{24}}(1-q)}{\eta(\tau)}\right|^2 \\ \nonumber &\times
\left[1-\frac{2\pi^2\tau_2\ell^2}{(12|\omega_1|)^2} (E_2(\tau)+12i \cot(\pi\tau) +c-13)(E_2(\btau)+12i \cot(\pi\btau) +c-13) + O(\ell^4) \right],
\end{align}
where we have used \eqref{eq:diff1} to evaluate the first correction. This provides a sharp prediction for the one-loop determinant of the massless graviton in  thermal AdS$_3$ / BTZ black hole with a radial cutoff.

In the bulk one tractable way to evaluate this correction to the one-loop determinant is by evaluating the quasinormal modes (QNMs) of the graviton of the BTZ black hole with a radial cut-off. The one-loop determinant can be written as a product over quasinormal modes \cite{Denef:2009kn}. The wave equation for the field can be solved exactly in the black hole background in terms of hypergeometric functions \cite{Birmingham:2001pj,Datta:2011za}. The ingoing boundary condition at the horizon is imposed as usual, but the Dirichlet boundary conditions are now imposed at the finite radial cut-off instead of at asymptotic infinity.
The spectrum of the QNMs can be solved perturbatively. This can then be used (along with suitable constraints for gauge-fixing) to verify whether \eqref{eq:vac} can be rederived from the bulk.

\section{Conclusions}
\label{sec:conc}

In ways more than one, the $\ttb$ deformation is special. Although it is an irrelevant deformation, it has several remarkable features and relates to starkly different kinds of physics. In this work, we have found yet another interesting feature of the deformed theories regarding the modular properties of the torus partition function. Based on the modular property, we derived a Cardy like formula for the asymptotic density of states which interpolates between a standard Cardy like behavior and a typical stringy Hagedorn like behavior. Modular invariance is a priori unexpected especially when irrelevant deformations are turned on, which typically demand the need of UV regulators. However, for the $\ttb$ deformation the UV cutoff does not appear manifestly in a number of quantities and partition function is still modular invariant. We also pointed out the existence of a BPS like sector in the spectrum and that the elliptic genus is invariant under $T\bar{T}$ deformation. Finally, we also commented on how the vacuum character changes under the $T\bar{T}$ deformation and a possible check from the holographic point of view.\par

As mentioned in the introduction, the $T\bar{T}$ deformation can be interpreted as coupling the field theory to a gravitational theory, either stochastic gravity \cite{Cardy:2018sdv} or the JT gravity \cite{Dubovsky:2018bmo}. It is an interesting question to understand the physical meaning of the modular properties from the point of view of quantum gravity. As a concrete starting point, the $T\bar{T}$ deformed free boson is nothing but the Nambu-Goto string where the deformation parameter $\ell^2$ plays the role of the string coupling constant $\alpha'$. Tuning up the $\ttb$ coupling is analogous to decreasing the string tension. The $\ell\to\infty$ limit then corresponds to the tensionless point where additional higher-spin symmetries emerge \cite{Gross}. In particular, the deformed spectrum \eqref{eq:deformed-energies} in this limit becomes $\E_n ^{(\infty)} \simeq P_n$ and the partition function is
\begin{align*}\label{eq:part-sp}
\Z(\omega_1,\omega_2,\ell\to \infty) \simeq   \sum_n e^{  i \tau |\omega_1| P_n }=Z_{\rm CFT}(\tau,\btau)\big|_{\btau=\tau},
\end{align*}
that is, the CFT partition function with $\tau=\btau$ is recovered. It will be intriguing to understand the implications of the above relation in greater detail.

Up to now the spectrum and the partition function (on non-trivial finite geometries \cite{Cardy:2018sdv,Dubovsky:2018bmo}) is reasonably understood although many questions remain open. From a purely mathematical standpoint, it would be interesting to construct classes of modular functions of the kind \eqref{eq:mod-trans-Z} and study them with a view towards relating them to the $\ttb$ deformed CFTs. It is also important to consider the deformations of other   observables such as correlation functions and entanglement entropies. A good starting point can be correlation functions of BPS operators identified in Section \ref{sec:BPS_sector}. These operators generically have non-vanishing OPE coefficients with non-BPS ones and therefore lead to correlation functions with non-trivial dependence on the $\ttb$ coupling. This would require an understanding how local operators of the CFT get deformed under the $T\bar{T}$ deformation. From the gravitational interpretation of the $T\bar{T}$ deformation, it might be quite subtle to define local operators. However, near the critical point, it may be possible to proceed using conformal perturbation theory, some steps in this direction were undertaken in \cite{Kraus:2018xrn}.\par

For computing quantities near critical points, it will be worthwhile to develop conformal perturbation theory for the $T\bar{T}$ deformation, both for the partition function and for other observables. The conformal perturbation theory involves divergent integrals and one needs to fix the prescription for doing conformal perturbation theory in finite geometries. One way to proceed is by using the results for the deformed free boson in Section \ref{sec:bosons} where both the deformed Lagrangian and partition functions are known explicitly. Once we understand how to perform integrals for the partition function, calculation of correlation functions can be potentially tackled.\par

Finally, for integrable CFTs, $T\bar{T}$ is only one of the members in the family of irrelevant operators that preserve integrability. It will be interesting to study how the spectrum and partition functions deform under these higher irrelevant operators \cite{Smirnov:2016lqw} and whether there are some interesting modular properties.

\section*{Acknowledgements}
We are grateful to John Cardy, Justin David, Matthias Gaberdiel and Per Kraus for valuable comments and suggestions on a draft of this paper. We thank Alex Maloney and John Cardy for suggesting the idea of finding modular properties from the diffusion equation which led to the subsection 2.3. It is  a pleasure to thank Ofer Aharony, Marco Billo, Michele Caselle, Lorenz Eberhardt, Amit Giveon, Monica Guica, David Kutasov and Rob Myers for fruitful discussions. This work is supported by the NCCR SwissMAP, funded by the Swiss National Science Foundation.

\appendix

\section{Details of some modular functions}
\label{sec:modularForm}
In this appendix, we give more details on the modular properties of various objects. We consider a generic modular transformation
\begin{align}
\gamma=\left(
         \begin{array}{cc}
           a & b \\
           c & d \\
         \end{array}
       \right)\in\text{SL}(2,\mathbb{Z})
\end{align}
It is easy to see that the imaginary part of the modular parameter $\tau_2$ transforms as
\begin{align}\label{eq:tau2-trans}
\tau_2\mapsto|c\tau+d|^{-2}\tau_2.
\end{align}
For the Dedekind eta function, we have
\begin{align}
\eta\left(\gamma\cdot\tau\right)=\epsilon(a,b,c,d)(c\tau+d)^{1/2}\eta(\tau)
\end{align}
where $\epsilon(a,b,c,d)$ is a phase factor that  cancels with a similar one in the anti-holomorphic part. The derivative of the Dedekind eta function is given by
\begin{align}
\label{eq:deta2}
\partial_{\tau}\eta(\tau)=\frac{\pi i}{12}\eta(\tau)E_2(\tau)
\end{align}
where $E_2(\tau)$ is the Eisenstein series.

The first few Eisenstein series are defined as
\begin{align}
E_2=1-24\sum_{m=1}^\infty\frac{m q^m}{1-q^m},\quad E_4=1+240\sum_{m=1}^\infty\frac{m^3 q^m}{1-q^m},\quad
E_6=1-504\sum_{m=1}^\infty\frac{m^5 q^m}{1-q^m}
\end{align}
where $q=e^{2\pi i\tau}$. Under the modular transformations, the Einsenstein series transform as
\begin{align}
E_2\left(\gamma\cdot\tau\right)=&(c\tau+d)^2 E_2(\tau)+\frac{6c}{\pi i}(c\tau+d),\\\nonumber
E_2\left(\gamma\cdot\bar{\tau}\right)=&(c\bar{\tau}+d)^2 E_2(\bar{\tau})-\frac{6c}{\pi i}(c\bar{\tau}+d)
\end{align}
and
\begin{align}
E_{2n}\left(\gamma\cdot{\tau}\right)=(c\tau+d)^{2n}E_{2n}(\tau),\qquad E_{2n}\left(\gamma\cdot\bar{\tau}\right)=(c\bar{\tau}+d)^{2n}E_{2n}(\bar{\tau})
\end{align}
for $n>1$. We notice that $E_2$ does not transform covariantly under modular transformation and is thus quasi-modular. This can be fixed by a shift, at the price of making it non-holomorphic (although it transforms holomorphically)
\begin{align}
\tilde{E}_2(\tau)=E_2(\tau)-\frac{3}{\pi\,\text{Im}\,\tau},\qquad \tilde{E}_2\left(\gamma\cdot\tau\right)=(c\tau+d)^2\tilde{E}_2(\tau).
\end{align}
It is exactly this combination that appears in the perturbative expansion of the free boson partition function in the main text.

The derivatives of the Eisentein series are given by the Ramanujan identities
\begin{equation}
\begin{aligned}
\label{eq:dtau}
\partial_{\tau}E_2(\tau)=&\,\frac{\pi i}{6}\left[E_2(\tau)^2-E_4(\tau)\right],\\
\partial_{\tau}E_4(\tau)=&\,\frac{2\pi i}{3}\left[E_2(\tau)E_4(\tau)-E_6(\tau)\right],\\
\partial_{\tau}E_6(\tau)=&\,\pi i\left[E_2(\tau)E_6(\tau)-E_4(\tau)^2\right].
\end{aligned}
\end{equation}
The above relations also allow us to express higher order derivatives of $\eta(\tau)$ in terms of $E_{2n}(\tau)$. For instance
\begin{align}\label{eq:d2eta}
\partial_{\tau}^2\eta(\tau)=\frac{\pi^2}{144}\left[2E_4(\tau)-3E_2(\tau)^2\right]\eta(\tau).
\end{align}
\section{Higher order partition functions}
\label{sec:diffOp}
\subsection*{General higher order corrections}
The higher order (from 5th to 10th order) expansions of the partition function in terms of the modular differential operators are as follows.
\begin{align}
&Z_5=\tfrac{4\tau_2^5}{15}\left(\mathscr{D}^{(4)}-\tfrac{10}{\tau_2^2}\mathscr{D}^{(3)}+\tfrac{127}{4\tau_2^4}\mathscr{D}^{(2)}
-\tfrac{36}{\tau_2^6}\mathscr{D}^{(1)}+\tfrac{45}{5\tau_2^8}\mathscr{D}^{(0)}  \right)Z_0,\\\nonumber
&Z_6=\tfrac{4\tau_2^6}{45}\left(\mathscr{D}^{(5)}-\tfrac{35}{2\tau_2^2}\mathscr{D}^{(4)}+\tfrac{427}{4\tau_2^4}\mathscr{D}^{(3)}
-\tfrac{2193}{8\tau_2^6}\mathscr{D}^{(2)}+\tfrac{1125}{4\tau_2^8}\mathscr{D}^{(1)}-\tfrac{675}{8\tau_2^{10}}\mathscr{D}^{(0)}\right)Z_0,\\\nonumber
&Z_7=\tfrac{8\tau_2^7}{315}\left(\mathscr{D}^{(6)}-\tfrac{28}{\tau_2^2}\mathscr{D}^{(5)}+\tfrac{581}{2\tau_2^4}\mathscr{D}^{(4)}
-\tfrac{1395}{\tau_2^6}\mathscr{D}^{(3)}+\tfrac{50553}{16\tau_2^8}\mathscr{D}^{(2)} \right.
\left. -\tfrac{6075}{2\tau_2^{10}}\mathscr{D}^{(1)}+\tfrac{14175}{16\tau_2^{12}}\right)Z_0,\\\nonumber
\end{align}\begin{align}
\nonumber
&Z_8=\tfrac{2\tau_2^8}{315}\left(\mathscr{D}^{(7)}-\tfrac{42}{\tau_2^2}\mathscr{D}^{(6)}+\tfrac{1365}{2\tau_2^4}\mathscr{D}^{(5)}  -\tfrac{5462}{\tau_2^6}\mathscr{D}^{(4)}+\tfrac{363033}{16\tau_2^8}\mathscr{D}^{(3)}\right.\\\nonumber
&\qquad\qquad\qquad\qquad\quad\left. -\tfrac{378171}{8\tau_2^{10}}\mathscr{D}^{(2)}+\tfrac{694575}{16\tau_2^{12}}\mathscr{D}^{(1)}-\tfrac{99225}{8\tau_2^{14}}\mathscr{D}^{(0)}\right)Z_0,\\\nonumber
&Z_9=\tfrac{4\tau_2^9}{2835}\left(\mathscr{D}^{(8)}-\tfrac{60}{\tau_2^2}\mathscr{D}^{(7)}+\tfrac{2877}{2\tau_2^4}\mathscr{D}^{(6)} -\tfrac{17747}{\tau_2^6}\mathscr{D}^{(5)}+\tfrac{1936089}{16\tau_2^8}\mathscr{D}^{(4)}\right.\\\nonumber
&\qquad\qquad\qquad\qquad\quad\left. -\tfrac{911367}{2\tau_2^{10}}\mathscr{D}^{(3)}+\tfrac{14308731}{108\tau_2^{12}}\mathscr{D}^{(2)}
-\tfrac{793800}{\tau_2^{14}}\mathscr{D}^{(1)}+\tfrac{893025}{4\tau_2^{16}}\mathscr{D}^{(0)}\right)Z_0,\\\nonumber
&Z_{10}=\tfrac{4\tau_2^{10}}{14175}\left(\mathscr{D}^{(9)}-\tfrac{165}{2\tau_2^2}\mathscr{D}^{(8)}+\tfrac{5577}{2\tau_2^4}\mathscr{D}^{(7)}  -\tfrac{200453}{4\tau_2^6}\mathscr{D}^{(6)}+\tfrac{8325009}{16\tau_2^8}\mathscr{D}^{(5)}\right.\\\nonumber
&\qquad\qquad\qquad\left.-\tfrac{101705877}{32\tau_2^{10}}\mathscr{D}^{(6)}+\tfrac{178354791}{16\tau_2^{12}}\mathscr{D}^{(3)}-
\tfrac{669294495}{32\tau_2^{14}}\mathscr{D}^{(2)}+\tfrac{72335025}{4\tau_2^{16}}\mathscr{D}^{(1)}-\tfrac{40186125}{8\tau_2^{18}}\right)Z_0.
\end{align}

\subsection*{Higher order corrections for the free boson}
We provide the 5th and 6th order for the free boson. Higher orders can be computed straightforwardly, but the results are too long to be displayed here
\begin{align*}
&Z_5=\tfrac{2\tau_2^5Z_0}{(12)^{10}}\left[-\tfrac{2\pi^{10}}{15}F_5-\tfrac{420\pi^8}{\tau_2^2}F_4-\tfrac{302400\pi^6}{\tau_2^4}F_3\right.
\left.-\tfrac{48988800\pi^4}{\tau_2^6}F_2-\tfrac{881798400\pi^2}{\tau_2^8}F_1+\tfrac{3174474240}{\tau_2^{10}}   \right],\\\nonumber
&Z_6=\tfrac{2\tau_2^6Z_0}{(12)^{12}}\left[\tfrac{2\pi^{12}}{45}F_6+\tfrac{1296\pi^{10}}{5\tau_2^2}F_5+\tfrac{408240\pi^8}{\tau_2^4}F_4
+\tfrac{195955200\pi^6}{\tau_2^6}F_3 +\tfrac{23808556800\pi^4}{\tau_2^8}F_2\right.\\\nonumber
&\qquad\qquad\qquad\left.+\tfrac{342843217920\pi^2}{\tau_2^{10}}F_1-\tfrac{1028529653760}{\tau_2^{12}}\right].
\end{align*}
Here the symbols $F_k$ are non-holomorphic modular forms of weight $4k$ and  are defined as
\begin{align}
&F_1=\left|\tilde{E}_2\right|^2,\\\nonumber
&F_2=\left|\tilde{E}_2^2-2E_4\right|^2,\\\nonumber
&F_3=\left|3\tilde{E}_2^3-18\tilde{E}_2E_4+16E_6\right|^2,\\\nonumber
&F_4=\left|15\tilde{E}_2^4-180\tilde{E}_2^2E_4-156 E_4^2+320\tilde{E}_2E_6\right|^2,\\\nonumber
&F_5=\left|105\tilde{E}_2(\tilde{E}_2^4-20\tilde{E}_2^2E_4-52E_4^2)+32(175\tilde{E}_2^2+58E_4)E_6\right|^2\\\nonumber
&F_6=\left|3(315\tilde{E}_2^6-9450\tilde{E}_2^4E_4-49140\tilde{E}_2E_4^2-3784E_4^3)+576\tilde{E}_2(175\tilde{E}_2^2+174E_4)E_6-14848E_6^2  \right|^2
\end{align}

\providecommand{\href}[2]{#2}

\end{document}